\nonstopmode

\newcommand{\im}{{\mathrm{Im}\,}}
\newcommand{\eV}{\U{eV}}
\newcommand{\Cal}[1]{{\cal #1}}
\newcommand{\ie}{i.e.{}}
\newcommand{\eg}{e.g.{}}
\newcommand{\U}[1]{\,{\rm{#1}}}
\newcommand{\I}[1]{_{\mathrm{#1}}}
\newcommand{\imag}{{\rm i}}
\newcommand{\euler}{\mathrm e}
\newcommand{\Int}{\int\limits}
\newcommand{\differential}{\>\mathrm d}
\newcommand{\XUV}{\textsc{xuv}}

\documentclass[pra,aps,twocolumn,longbibliography,showpacs,preprintnumbers]{revtex4-1}
\usepackage{bm,bbm,amsmath,amssymb,subeqnarray,graphicx}
\usepackage[nativepdf,bookmarks,bookmarksopen,bookmarksnumbered,raiselinks,breaklinks,%
pdftitle={Ramsey method for Auger-electron interference induced by an attosecond twin pulse},%
pdfauthor={Christian Buth, Kenneth J. Schafer},%
pdfsubject={Atomic physics},%
pdfkeywords={attosecond science, XUV, photoabsorption, Auger decay, %
interferometry, krypton, attosecond twin pulse, Norman F. Ramsey, %
method of separated oscillatory fields, quantum control, decoherence meter, %
essential-states model}]{hyperref}

\begin{document}
\title{Ramsey method for Auger-electron interference induced by an
attosecond twin pulse}
\author{Christian Buth}
\thanks{Corresponding author.  Present address: Theoretische Chemie,
Physikalisch-Chemisches Institut, Ruprecht-Karls-Universit\"at Heidelberg,
Im Neuenheimer Feld~229, 69120~Heidelberg, Germany.
Electronic mail}
\email{christian.buth@web.de}
\author{Kenneth J.~Schafer}
\affiliation{The PULSE Institute for Ultrafast Energy Science,
SLAC National Accelerator Laboratory, Menlo Park, California 94025, USA}
\affiliation{Department of Physics and Astronomy, Louisiana State
University, Baton Rouge, Louisiana~70803, USA}
\date{Received 17 February 2010; revised 11 November 2014;
      published 18 February 2015}

\begin{abstract}
We examine the archetype of an interference experiment for Auger electrons:
two electron wave packets are launched by inner-shell ionizing a krypton
atom using two attosecond light pulses with a variable time delay.
This setting is an attosecond realization of the Ramsey method of separated
oscillatory fields.
Interference of the two ejected Auger-electron wave packets
is predicted, indicating that the coherence between the two pulses is passed
to the Auger electrons.
For the detection of the interference pattern an accurate
coincidence measurement of photo- and Auger electrons is necessary.
The method allows one to control inner-shell electron dynamics
on an attosecond timescale and represents a sensitive indicator for
decoherence.
\end{abstract}

%
%
%
%
%
%
%

\preprint{arXiv:1002.3336}
\maketitle

\renewcommand{\onlinecite}[1]{\cite{#1}}

\section{Introduction}

Ramsey's method of separated oscillatory
fields~\cite{Ramsey:TM-80,Ekspong:NL-93}
represents a paradigm of precision measurement for various physical
quantities.
In its original conception for the measurement of nuclear magnetic moments,
the scheme uses two coherent radiation fields, which are separated by
a field-free propagation interval.
The signature of the coherent interaction is the appearance of
interference fringes when the physical quantity under consideration
is measured at the exit of this experimental setup.
Since Ramsey's seminal studies, his method has been extended and modified
extensively, \eg, by considering multiple fields with varying phase
and amplitude and by applying it to masers and lasers~\cite{Ramsey:TM-80}.
The method of separated oscillatory fields is an
interferometric approach which has the advantage over pump-probe
schemes in that it does not depend on an intense pump pulse~\cite{Krausz:AP-09}.

\begin{figure}
  \begin{center}
    \includegraphics[clip,width=\hsize]{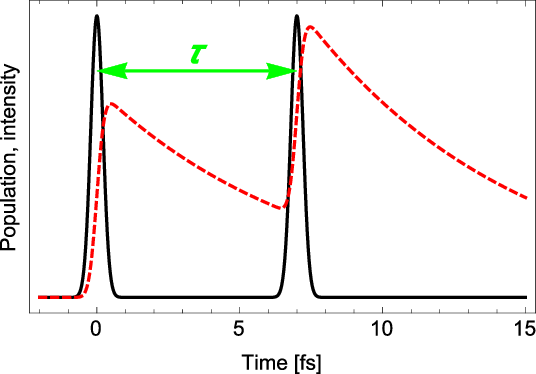}
    \caption{(Color online) The \XUV{}~intensity~$I\I{X}(t)$ (solid, black) and
             krypton~$3d$ hole population~$\varrho(t)$~(dashed, red)
             [Eq.~(\ref{eq:ratepop})] for two attosecond pulses
             ($\tau\I{X} = 500 \U{as}$) separated by a variable delay
             of~$\tau$.}
    \label{fig:TwoAttoPulses}
  \end{center}
\end{figure}

In this article, we propose a Ramsey scheme for attosecond science
assuming two coherent pulses with a full width at half maximum~(FWHM)
duration of~$\tau\I{X} = 500 \U{as}$ each, which are separated by a
variable delay of~$\tau$ [Fig.~\ref{fig:TwoAttoPulses}]~\cite{Buth:AD-09}.
With an essential-states model~\cite{Smirnova:QC-03,Buth:TA-09}, we
investigate the situation where a twin pulse ionizes the
$3d$~shell of krypton atoms;
the $3d$~holes subsequently decay in terms of
an $M_{4,5} N_1 N_{2,3}$~Auger process.
In fact, one of the first applications in attosecond science was the
determination of the time constant of this Auger decay channel---a
well-known datum from frequency-domain spectroscopy,~$\tau_{3d} =
7.5 \U{fs}$---with a single attosecond pulse in the presence of an
optical streaking laser~\cite{Drescher:TR-02}.
Unlike this first investigation, our proposal of an attosecond Ramsey method
[Fig.~\ref{fig:TwoAttoPulses}] represents a fundamental experiment
which is only feasible with attosecond science and has no
frequency-domain equivalent in the sense that the attosecond twin pulse
is crucial for its realization.
Nonetheless, the spectra in this paper are shown in the frequency domain.
Furthermore, Auger decay~\cite{Auger:SR-23} is a pure
manifestation of electron correlations:
it cannot be understood in terms of an effective one-electron model.

The twin pulse shown in Fig.~\ref{fig:TwoAttoPulses} induces two outgoing
photoelectron and Auger-electron waves.
Interference between two photoelectron wave packets
was examined in Ref.~\onlinecite{Wollenhaupt:IU-02}.
In that study, $5p$~Rydberg electrons of potassium atoms were
subjected to two identical time-delayed laser pulses with
a FWHM duration of~$30 \U{fs}$ each at~$790 \U{nm}$~wavelength
and the resulting interference pattern in
the photoelectron spectrum was analyzed.
In contrast to the experiment in Ref.~\onlinecite{Wollenhaupt:IU-02},
we ask whether the coherence of the light is also transferred to the
Auger electrons and what kind of Auger-electron spectrum can we expect
if it is?
Clearly, for a time delay between the two pulses of~$\tau = 0$
and of~$\tau \to \infty$, we observe no interference fringes.
What happens in between the two limiting cases?
Clearly, our proposal of an attosecond Ramsey scheme represents an
important experimental test of our understanding of Auger decay.
Furthermore, the scheme also represents a versatile tool for measurements.
It enables one to precisely determine the position of Auger lines and
it is a measure of coherence.
Such an experiment would also address the following questions:
how much decoherence is caused by the Auger process and what
is the coherence time?
Is our understanding of Auger decay complete?
Atomic units are used throughout unless stated otherwise.

\section{Theory}

The simplest way to describe Auger decay is shown in
Fig.~\ref{fig:TwoAttoPulses}.
Here, a rate-equation model which
is used to determine the probability to find a $3d$~hole at time~$t$ in
krypton~$\varrho(t)$~\cite{Drescher:TR-02}.
It is given by the convolution of exponential Auger decay
with a width of~$\Gamma = 88 \U{meV}$~\cite{Jurvansuu:IL-01,Drescher:TR-02}
with the \XUV{}~intensity
\begin{equation}
  \label{eq:ratepop}
  \varrho(t) = \dfrac{\sigma}{\omega\I{X}} \Int_{-\infty}^t
    I\I{X}(t') \; \euler^{-\Gamma (t - t')} \differential t' \; .
\end{equation}
The absorption cross section~$\sigma$ is taken to be constant
over the bandwidth of the \XUV{}~pulse with a central angular frequency
of~$\omega\I{X}$ and an intensity of~$I\I{X}(t')$ at time~$t'$.
The model does not honor the phase relationship between the
two ejected Auger-electron wave packets and thus does not
describe interference effects~\cite{Smirnova:QC-03}.

To treat the quantum-mechanical phases correctly, we use an \emph{ab initio}
formalism for the quantum dynamics of Auger decay of
atoms which are inner-shell ionized by extreme
ultraviolet~(\XUV)~light~\cite{Buth:TA-09}.
The attosecond pulses of present-day light sources have a low peak
intensity and their interaction may be described perturbatively
as a one-photon process~\cite{Krausz:AP-09}.
The quantum dynamics of the inner-shell hole creation with subsequent
Auger decay is given by equations of motion which we simplify here
in terms of an essential-states model~\cite{Smirnova:QC-03,Buth:TA-09}.
Our theory yields the probability density
amplitude~$\bar c\I{A}^{\vec k\I{P} \, \vec k\I{A}}(\tau)$
to find a photoelectron
with momentum~$\vec k\I{P}$ in coincidence with an Auger electron
with momentum~$\vec k\I{A}$ for a delay of~$\tau$ between
the two pulses in Fig.~\ref{fig:TwoAttoPulses}.
The probability density amplitude is adapted for no laser dressing
from Eqs.~(61) and (62) of Ref.~\onlinecite{Buth:TA-09};
it reads
\begin{equation}
  \label{eq:lasdreauampre}
  \bar c\I{A}^{\vec k\I{P} \, \vec k\I{A}}(\tau) = \frac{\imag}{2} \>
    \bar d(\vec k\I{P}) \> \bar v(\vec k\I{A}) \; S\Bigl(\tau,
    \frac{\vec k\I{P}^2}{2}, \frac{\vec k\I{A}^2}{2} \Bigr) \; ,
\end{equation}
with the rms dipole and rms Auger decay matrix elements~$\bar d(\vec k\I{P})$
and $\bar v(\vec k\I{A})$, respectively.
The lineshape function in Eq.~(\ref{eq:lasdreauampre}) is
\begin{equation}
  \label{eq:lineshape}
  S(\tau, \omega\I{P}, \omega\I{A}) = \frac{\tilde \varepsilon\I{X}
    (\tau, \omega\I{P} + \omega\I{A} - \Omega\I{P} - \Omega\I{A})}
    {\omega\I{A} - \Omega\I{A} - \Delta\I{R} + \imag \frac{\Gamma}{2}} \; ;
\end{equation}
it depends only on the absolute values of the momenta~$k\I{P} = |\vec k\I{P}|$
and $k\I{A} = |\vec k\I{A}|$.
Furthermore, it contains the nominal photoelectron and Auger
electron energies, which are in our case~$\Omega\I{P} = 20 \eV$
and $\Omega\I{A} = 40 \eV$, respectively~\cite{Buth:TA-09}.
In Eq.~(\ref{eq:lineshape}), $\Delta\I{R}$~is the second-order energy
shift and $\Gamma$~is the Auger decay width~\cite{Buth:TA-09}.
The spectral envelope of the \XUV{}~light for a twin pulse is given by
\begin{equation}
  \label{eq:spectraltwin}
  \tilde \varepsilon\I{X}(\tau, \omega) = \sqrt{\frac{\pi}{2\log 2}} \;
    \varepsilon\I{X0} \, \tau\I{X} \>
    \euler^{-\frac{\omega^2 \tau\I{X}^2}{8 \log 2}} \>
    ( 1 + \euler^{\imag \omega  \tau} ) \; ,
\end{equation}
with the peak electric field strength~$\varepsilon\I{X0}$.

\section{Results and Discussion}

\begin{figure}
  \begin{center}
    \includegraphics[clip,width=\hsize]{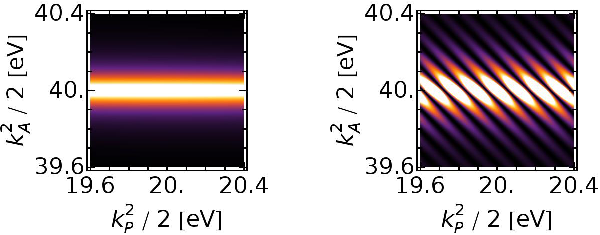}
    \caption{(Color online) Probability density~$|\bar c\I{A}^{\vec k\I{P} \,
             \vec k\I{A}}(\tau)|^2$ [Eq.~(\ref{eq:lasdreauampre})]
             to find a photoelectron with~$k\I{P}$ and an Auger electron
             with~$k\I{A}$.
             We average the photoelectron over the full solid angle and
             view the Auger electrons along the $z$~axis (linear
             \XUV{}~polarization axis).
             The left panel is for a delay of~$\tau = 0$ and the right panel is
             for~$\tau = 5 \, \tau_{3d}$.
             The color scale is linear.}
    \label{fig:unnorm2d}
  \end{center}
\end{figure}

The main result of this study, the probability
density~$|\bar c\I{A}^{\vec k\I{P} \, \vec k\I{A}}(\tau)|^2$
[Eq.~(\ref{eq:lasdreauampre})], is displayed in Fig.~\ref{fig:unnorm2d}
for both no time delay between the two pulses of Fig.~\ref{fig:TwoAttoPulses},
\ie, a single pulse, and a time delay of~$\tau = 5 \, \tau_{3d}$.
The second choice for~$\tau$ is somewhat arbitrary;
the value of~$5 \, \tau_{3d}$ is high enough to cause significant
structure in the right panel of Fig.~\ref{fig:unnorm2d}.
This indicates interference effects that we would like to analyze
in the following.
The shape of the plots in Fig.~\ref{fig:unnorm2d} is determined by the
absolute square of the line shape function~(\ref{eq:lineshape}) in
a nontrivial way.
For~$\tau = 0$, horizontally, along the $k\I{P}^2 / 2$~coordinate,
the width of the line profile is determined by the FWHM
of~$|\tilde \varepsilon\I{X}(0, \omega)|^2$, which in our case
%
%
is~$3.7 \eV$.
Vertically, along the $k\I{A}^2 / 2$~coordinate, the extension is defined
by the Auger decay width of~$88 \U{meV}$~\cite{Jurvansuu:IL-01,Drescher:TR-02}.
In the case of~$\tau = 5 \, \tau_{3d}$, we have a more involved dependence;
overall, the contour has the shape of a skewed hyperbola with respect
to~$k\I{A}$ caused by the denominator squared in Eq.~(\ref{eq:lineshape}).
For a deeper understanding of Fig.~\ref{fig:unnorm2d}, we realize that the
emission of an Auger electron is in fact a correlated two-electron process
of photoionization and electronic decay.
For such a process, we can exploit the energy
balance~\cite{Smirnova:EC-05,Krausz:AP-09}:
\begin{equation}
  \label{eq:energybalance}
  \frac{\vec k\I{P}^2}{2} + \frac{\vec k\I{A}^2}{2} = \omega\I{X} - I^{++} \; .
\end{equation}
Here, $I^{++}$~is the double-ionization potential of the neutral atom
for producing the dicationic final state.
The balance~(\ref{eq:energybalance}) manifests in the argument
of~$\tilde \varepsilon\I{X}$ in Eq.~(\ref{eq:lineshape}) and,
consequently, it is reflected by the diagonal lines in the right
panel of Fig.~\ref{fig:unnorm2d}.
Relation~(\ref{eq:energybalance}) is only exact for
monochromatic \XUV{}~light with photon energy~$\omega\I{X}$, \ie, a
continuous-wave source.

\begin{figure}
  \begin{center}
    \includegraphics[clip,width=\hsize]{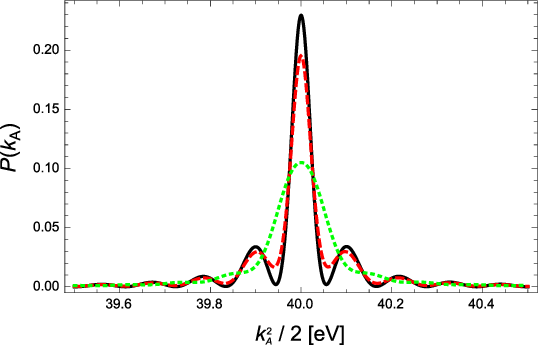}
    \caption{(Color online) Auger-electron spectrum viewed along the $z$~axis
             for an attosecond twin pulse with a separation of~$\tau
             = 5 \, \tau_{3d}$.
             The interference fringes are diminished for a decreasing
             accuracy of the photoelectron measurement;
             the energy uncertainties are: $\pm 0.04 \eV$~(solid, black),
             $\pm 0.1 \eV$~(dashed, red), and $\pm 0.4 \eV$~(dotted, green).}
    \label{fig:AugerDelay}
  \end{center}
\end{figure}

In a typical Auger-electron-spectroscopy experiment, the photoelectron
is not observed.
Hence we need to integrate the probability
density~(\ref{eq:lasdreauampre}) over the unobserved degrees of
freedom, which is in this case the photoelectron momentum~$\Int_{\mathbb{R}^3}
|\bar c\I{A}^{\vec k\I{P} \, \vec k\I{A}}(\tau)|^2 \differential^3 k\I{P}$.
For each Auger-electron momentum, this implies an integration along
a horizontal line in Fig.~\ref{fig:unnorm2d}.
Following such a path in the right panel visually, we see that we
average over many fringes with different energies which leads to a
washing out of the interference pattern.
Indeed, the resulting Auger-electron spectrum for an unknown photoelectron
momentum exhibits no noticeable fringes;
it resembles closely the dotted, green curve in Fig.~\ref{fig:AugerDelay}.

Inspecting the lineshape function~(\ref{eq:lineshape}), we find that
the \XUV{}~envelope is imprinted on the Auger electrons
due to correlations between photoelectrons and Auger
electrons.
The finding that integrating over the photoelectron diminishes interference
effects conforms to the general fact that summing over unobserved degrees of
freedom generally comes with a loss of coherence.
Consequently, to preserve the coherence of the Auger decay process,
we need to take the momentum of the photoelectron into account.

The other way around, however, does not hold true: it is \emph{not}
required to observe the Auger electron to see interference
fringes in the photoelectron spectrum~\cite{Wollenhaupt:IU-02}.
To see why this is so, we derive the probability density to observe a
photoelectron.
Within our formalism~\cite{Buth:TA-09} it is given by
\begin{equation}
  \label{eq:pelspec}
  \tilde{\Cal P}\I{P}(\tau, \vec k\I{P}) = \dfrac{|\bar d(\vec k\I{P})|^2}
    {2 \pi} \Int_{-\infty}^{\infty} \im \biggl [
    \dfrac{ |\tilde \varepsilon\I{X}(\tau, \omega) |^2}
    {\frac{\vec k\I{P}^2}{2} - \Omega\I{P} + \Delta\I{R} - \omega
    - \imag \frac{\Gamma}{2}} \biggr ] \differential \omega \; .
\end{equation}
Our analysis has revealed the interconnection of the photoionization
and the subsequent Auger decay;
the dependence of Eq.~(\ref{eq:pelspec}) on the Auger decay width
represents the reciprocal connection.
This can be understood as follows: the interference of the photoelectrons is
caused by the envelop~$\tilde \varepsilon\I{X}$ of the \XUV{}~light.
The only impact of Auger decay on the photoelectron
is due to the filling of the created hole which leads to a
line broadening that leads for large Auger widths to a
washing out of interference fringes.

\begin{figure}
  \begin{center}
    \includegraphics[clip,width=\hsize]{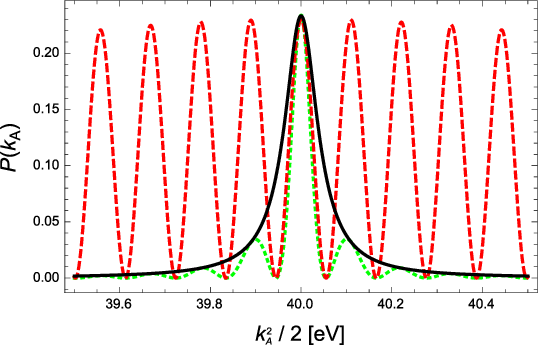}
    \caption{(Color online) Interference of the Auger-electron waves~(dotted,
             green) from an attosecond twin pulse for a precisely known
             photoelectron momentum~$k\I{P}^2 / 2 = \Omega\I{P}$.
             The interference pattern is decomposed into the scaled
             lineshape function~(solid, black) and the scaled \XUV{}~spectral
             envelope square~(dashed, red).}
    \label{fig:AugerInterf}
  \end{center}
\end{figure}

To have a chance of observing interference between Auger-electron waves,
we need to preserve the coherence from the two \XUV{}~pulses.
To accomplish this goal, we recall the energy balance in
Eq.~(\ref{eq:energybalance}).
It implies that if we measure the photoelectron momentum~$k\I{P}$
with a certain precision, this defines the uncertainty in
the Auger-electron momentum~$k\I{A}$.
In other words, if we restrict the allowed photoelectron momenta to
a narrow range, the destructive interference of Auger waves should be
reduced significantly.
We integrate~$|\bar c\I{A}^{\vec k\I{P} \, \vec k\I{A}}(\tau)|^2$
over the full solid angle and a specified  photoelectron momentum
range~$\Delta k$.
This yields the probability~$P_{\mathrm{PA}, \Delta k}
(\tau, k\I{P},k\I{A})$ to observe an Auger electron
with~$k\I{A}$---we look along the
$z$~axis---for a photoelectron in the full solid angle
with a momentum magnitude in the range
of~$[ \max\{0, k\I{P} - \Delta k \} ; k\I{P} + \Delta k ]$.
By integrating Eq.~(\ref{eq:pelspec}) over the same angular and momentum range,
we find the normalized probability distribution of the
photoelectrons~$P_{\mathrm P, \Delta k}(\tau, k\I{P})$~\cite{Buth:TA-09};
if we observe the photoelectron to lie in a chosen range,
then the conditional probability to find an Auger electron with a specific
momentum along the $z$~axis~$P_{\mathrm A, k\I{P}, \Delta k}(\tau, k\I{A})$
follows from Bayes law~\cite{Boas:MM-83}:
\begin{equation}
  \label{eq:BayesLaw}
  P_{\mathrm A, k\I{P}, \Delta k}(\tau, k\I{A}) =
    \frac{P_{\mathrm{PA}, \Delta k}(\tau, k\I{P},k\I{A})}
    {P_{\mathrm P, \Delta k}(\tau, k\I{P})} \; .
\end{equation}

In Fig.~\ref{fig:AugerDelay}, we investigate the Auger-electron spectrum
from Eq.~(\ref{eq:BayesLaw}).
For a very accurate measurement of the photoelectron momentum
(small uncertainty in~$\Delta k$), we find significant
interference fringes.
The interference effects are diminished with growing~$\Delta k$, \ie,
%
%
we average over Auger waves.
The first maximum off the main peak in the curves
of Fig.~\ref{fig:AugerDelay} moves to higher energies
with increasing~$\Delta k$.

In order to analyze the origin of the interference effects,
we assume an exactly known photoelectron momentum magnitude
and an exact detection of the Auger electrons along the $z$~axis.
In other words, we view the plots of Fig.~\ref{fig:unnorm2d}
along a vertical line and normalize it to the
peak of the photoelectron spectrum.
In Fig.~\ref{fig:AugerInterf}, we show a lineout of the right panel
of Fig.~\ref{fig:unnorm2d} for~$k\I{P}^2 / 2 = \Omega\I{P}$.
The Auger-electron spectrum is decomposed
into a lineshape function and the spectral \XUV{}~pulse
envelope~(\ref{eq:spectraltwin}) square.
The linewidth only depends on the time delay~$\tau$ as it
should in the Ramsey method~\cite{Ramsey:TM-80,Ekspong:NL-93}
and the interference fringes in the \XUV{}~field
envelope get thinner for increasing delay between the two pulses.
The spectral width between the first minimum on the left and the first
minimum on the right of the central peak of the fringes
%
%
is~$2 \pi / \tau = 0.1 \eV$ for~$\tau = 5 \, \tau_{3d}$.

\begin{figure}
  \begin{center}
    \includegraphics[clip,width=\hsize]{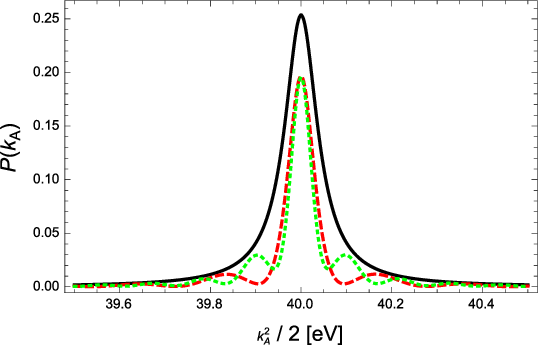}
    \caption{(Color online) Interference of the Auger-electron waves
             for different time delays~$\tau$ between the two attosecond
             pulses in Fig.~\ref{fig:TwoAttoPulses}: $\tau = 0$ (solid, black),
             $\tau = 3 \, \tau_{3d}$ (dashed, red), and $\tau = 5 \, \tau_{3d}$
             (dotted, green) for a photoelectron momentum which is known with
             an uncertainty of~$(\Delta k)^2/2 = 10^{-5} \eV$.}
    \label{fig:delay}
  \end{center}
\end{figure}

For a system with decoherence, we assume that one will find a similar
behavior of the interference pattern as for an inaccurately measured
photoelectron momentum.
In that case, in addition to averaging over waves with different
wavelengths, also a jitter in the phase relation
due to a coupling to other degrees of freedom
in the system suppresses the interference fringes.
Depending on the nature of the decoherence, a model of its
impact on the signal can be made.
Given a prediction of the signal, the observed Auger-electron interference
pattern and its change, when the time delay~$\tau$ in
Fig.~\ref{fig:TwoAttoPulses} is varied, can be used
to identify and measure the amount of decoherence in a system.
The impact of a variation of~$\tau$ for our perfectly coherent
case is revealed in Fig.~\ref{fig:delay}.
The curves resemble the ones in Fig.~3 of Ref.~\onlinecite{Wollenhaupt:IU-02}.
Thus an immediate application of the scheme discussed here is that
it is a coherence meter.

The experimental investigation that we propose is challenging because a
coincident detection of two electrons is necessary.
In a recent experimental study~\cite{Prumper:AP-08}, electrons from the
$KVV$~Auger decay of a C$\,1s$~vacancy in a CO~molecule were measured
in coincidence with the angular distribution of O$^+$~fragments.
This technique can be used also in our case:
the detection of the energy of the Auger electrons ejected along
a specific direction in coincidence with the measurement of the
momentum of the krypton ionic remnant offers an indirect pathway
for a coincidence experiment for photo- and Auger electrons.

\section{Conclusion}

In this article, we have proposed a fundamental experiment for studying the
attosecond science equivalent of the Ramsey method of separated oscillatory
fields: to what degree can we control an ultrafast electronic process in the
time domain~\cite{Krausz:AP-09}?
The setting can be used as a meter for decoherence in a system
and offers interesting perspectives when used with intense \XUV~light
tuned to an atomic resonance~\cite{Demekhin:IN-12,Adams:QO-13,Picon:OC-13}
where emission of a photoelectron can be avoided.

\begin{acknowledgments}
This work was supported by the National Science Foundation Grants
No.~PHY-0701372 and No.~PHY-0449235.
\end{acknowledgments}

\end{document}